# Scaling stellar jets to the laboratory:

# the power of simulations

Chantal Stehlé [1], Andrea Ciardi [1,2],

Jean-Philippe Colombier [1,3], Matthias González [4], Thierry Lanz [1,5], Alberto Marocchino [6],

Michaela Kozlova [7], Bedrich Rus [7]

[1] LERMA, Observatoire de Paris, CNRS et UPMC, 5 place J. Janssen, 92195 Meudon, France

[2] LPP, 10/12 avenue de l'Europe, 78140 Vélizy, France

[3] Laboratoire Hubert Curien, UMR CNRS 5516, 18 Rue du Professeur Benoît Laura, 42000 Saint-Etienne, France

[4] Instituto de Fusión Nuclear, Universidad Politécnica de Madrid, Madrid, Spain

[5] Department of Astronomy, University of Maryland, College Park, MD 20742, USA

[6] The Blackett Laboratory, Imperial College, Prince Consort Road, London SW7 2AZ, UK

[7] Department of X Ray Lasers, Institute of Physics PALS Center, Prague 8, Czech Republic

**Corresponding author** : Chantal Stehlé, Observatoire de Paris, LERMA, 5 Place Jules Janssen, 92195 Meudon France +33 1 45 07 74 16  mail : chantal.stehle@obspm.fr

**Short title** : Scaling stellar jets to the laboratory

**Number of manuscript pages, including figures  :30**

**Number of tables  : 1**

**Number of figures : 9**

# Scaling stellar jets to the laboratory: the power of simulations

**ABSTRACT**

Advances in laser and Z-pinch technology, coupled with the development of plasma diagnostics and the availability of high-performance computers, have recently stimulated the growth of high-energy density laboratory astrophysics. In particular a number of experiments have been designed to study radiative shocks and jets with the aim of shedding new light on physical processes linked to the ejection and accretion of mass by newly born stars.

Although general scaling laws are a powerful tools to link laboratory experiments with astrophysical plasmas, the phenomena modelled are often too complicated for simple scaling to remain relevant. Nevertheless, the experiments can still give important insights into the physics of astrophysical systems and can be used to provide the basic experimental validation of numerical simulations in regimes of interest to astrophysics.

We will illustrate the possible links between laboratory experiments, numerical simulations and astrophysics in the context of stellar jets. First we will discuss the propagation of stellar jets in a cross-moving interstellar medium and the scaling to Z-pinch produced jets. Our second example focuses on slab-jets produced at the PALS (Prague Asterix Laser System) laser installation and their practical applications to astrophysics. Finally, we illustrate the limitations of scaling for radiative shocks, which are found at the head of the most rapid stellar jets.

# 1- Introduction

In their infancy, stars originating from the collapse of dense interstellar clouds continue to accrete from the residual disk and envelope, while at the same time expelling powerful collimated jets. These jets, and the shocks that are produced as they propagate through the surrounding Interstellar Medium (ISM), can extend over distances of several parsecs (1 parsec ~ 2 $10^5$ AU ~ 3 $10^{16}$ m) and are generally observed over a large spectral range (Snell et al. 1980, Mundt & Fried 1983; Bieging et al 1984). Jets often appear a sequence of bright emission features that are consistent with internal shock produced by small velocity variations along the flow [Reipurth et al. 1992, 2002]. However the mechanism responsible for such variability is not clear. For instance it is not known whether this knotty structure, as traced by the so-called Herbig-Haro objects, is a consequence for example of instabilities occurring during the jet propagation, or if it is due to an intrinsic variability of the jet ejection process. Whereas jet launching is linked to magnetic fields anchored to a disk, and which are responsible for accelerating and collimating the jet near the star, magnetic fields at larger distances from the source seem to be too small to have any influence on the jet dynamics [Hartigan et al. 2007]. This can thus be accurately described by hydrodynamics, and in this context we discuss experiments investing the physics of astrophysical jets. The article is organised as follows. To be able to link laboratory to astrophysical flows, certain scaling requirements need to be satisfied [Ryutov et al. 1999, 2000,2001], and the basic ideas will be reviewed in section §2. The link from laboratory to astrophysics and the interpretation of the results are performed with appropriate multidimensional codes, which are described in §3. In section §4, using the case of curved jets produced on Z-pinch experiments performed on the MAGPIE facility (Lebedev et al. 2004, Ciardi et al 2008), we discuss how scaling may be used to understand the dynamics of the curved jets from Young Stellar Objects (YSOs). In §5 we will present preliminary interpretation of new experiments on the PALS laser facility, which use innovative laser focusing geometries to generate slab-jets in the relevant astrophysical parameter range. The question of shocks in stellar jets, and more specifically of radiative shocks is discussed in §6.

# 2- Scaling

The issues related to the scaling astrophysical flows to the laboratory were addressed by Ryutov et al., 1999, 2000, 2001 in the framework of ideal (negligible viscosity, thermal conductivity, and resistivity) compressible magneto-hydrodynamics. Here we shall only be concerned with hydrodynamic scaling between laboratory (L) and astrophysical (A) flows, which have typical lengths L*, velocity V* density ρ∗, and temperature T*, which relies on the invariance of the inviscid hydrodynamic equations to the transformation:

$$r_L = a r_A \; ; \; r_L = b r_A \; ; \; P_L = c P_A \, , \text{ thus } t_L = a(b/c)^{1/2} t_A \; ; \; u_L = (c/b)^{1/2} u_A \qquad (1)$$

where *r* denotes the position, t the time, and the other quantities ρ (density), P (pressure) and *u* velocity, are functions of position and time. The constants a, b, c are determined by the initial conditions which have to be geometrically similar for both the astrophysical and laboratory systems. Under these conditions, and provided that the so-called Euler number, $Eu = V^* (\rho^* / P^*)^{1/2}$, is the same for the two systems, the dynamics of these flows will be indistinguishable up the time-scale transformation. Although producing geometrically similar initial conditions is rarely possible, experimental flows and shocks in the correct regime represent a unique tool to study astrophysically relevant processes in the laboratory. The "correct regime" assumption here means that the laboratory flows produced can be described by the inviscid Euler equations so that Peclet number $Pe = L^* V^* / \chi$ ($\chi$ being the heat diffusion coefficients) and Reynolds number $Re = L^* V^* / \nu$, $\nu$ being the viscosity diffusion coefficient) are much larger than unity. Typical physical conditions for astrophysical and laboratory jets are given in Table 1.

Obtaining an Exact similarity is even more difficult in the presence of a radiation field (Ryutov et al. 1999, 2001; Castor 2007). The nature of the coupling between radiation and hydrodynamics is determined by the value of the optical mean free path compared to typical lengths, or equivalently by the optical depth $\tau = \kappa L^*$ where $\kappa$ is the medium's opacity. Jets from YSOs are assumed to be optically thin ($\tau \ll 1$), with the exception of jets propagating with relatively larger velocities (> 200 km s$^{-1}$) and which can generate radiative shocks with a developed precursor (Raga 1999).

Scaling remains theoretically feasible for optically thin plasmas, where it may be included as a cooling (sink) term in the energy equation. If the cooling terms vary as $\rho^\alpha P^\beta$ with the same power exponents for the two systems, then another condition between P* and ρ∗ is required for scaling (see Eq. 20, Ryutov et al. 2001). In practice, such precise scaling remains mostly intractable and the strategy consists then in checking that both systems are in a similar regime expressed by the dimensionless parameter $\chi_{cool}$ = V* $\tau_c$ */ L* = $L_{cool}$ / L* , where $\tau_c$ is the ratio of the gas thermal energy to the radiated energy per unit time and $L_{cool}$ is the cooling length. Adiabatic jets have $\chi > 1$, while radiatively cooled flows, such as YSO jets, have $\chi < 1$.

Scaling fails for the radiative bow shock produced at the head of high-velocity jets. In this case, the unshocked gas absorbs the UV photons emitted by the shock, where temperatures reach ~ 3 $10^5$ K, and a radiative precursor is generated. The Euler equations need then to be complemented with the equations for radiation transport, which are inherently non-local, and which are coupled to the population equations for all ionic and excitation stages of the chemical species in the gas.

## 3- Numerical modelling

While scaling helps developing experiments in the appropriate regime, to understand the limitations and relevance of laboratory flows to astrophysical models, we need to rely on numerical simulations. In this work we present simulations of experiments performed on both laser and z-pinch facilities. Astrophysical and laboratory jet simulations are performed with GORGON, a three-dimensional (3D) resistive magneto-hydrodynamic code (Chittenden et al. 2004; Ciardi et al. 2007). In the case of laboratory plasmas, local thermodynamic equilibrium (LTE) is assumed; the average ionization is calculated by a Thomas-Fermi model. The ion and electron energies are solved separately and include electron and ion thermal conductions, and optically-thin radiation losses. To model astrophysical flows, GORGON includes the time-dependent ionization of hydrogen by taking into account the recombination and collisional ionization and uses rate coefficients as tabulated in Raga et al. (2007). For temperatures above 15000 K, cooling is implemented by a function appropriate for interstellar gas composition (Dalgarno & McCray 1972). For temperatures below 15000 K, cooling is calculated by including the collisional excitation and ionization of hydrogen, radiative recombination of hydrogen, and the collisional excitation of O I and O II. The populations of O I and O II are assumed to follow closely those of hydrogen because of charge exchange (Hartigan & Raymond 1993).

The simulations related to the experiments on radiative shocks have been performed with HERACLES, a 3D radiation hydrodynamics code, which solves the Euler equations for hydrodynamics coupled with the moment equations of the radiative transfer equation. The M1 model allows us to bridge the transport and diffusive limits (González et al. 2007). The LTE approximation is assumed. The equation of state is computed with the OPA_CS code, which is based on the hydrogenic model described in Michaut et al. (2004). Opacities are calculated using the STA code (Bar Shalom et al. 1989) and are used for the grey radiation transport (González et al. 2006, 2009).

**4- Jet deflection**

A number of bipolar Herbig-Haro jets exhibit a distinguishing C-shape morphology indicative of a steady bending (Bally & Reipurth 2001). This curvature is attributed either to the motion of the jet source relative to the local ISM or to the presence of an extended flow, such as a wind from a nearby star. In general this give rise to an "effective" transverse wind that curves the jet, with expected "cross-wind" velocities varying from a few km s$^{-1}$ to few tens of km s$^{-1}$ (Jones & Herbig 1979; Salas et al. 1998; Bally & Reipurth 2001). The effect of jet bending by a lateral flow has been studied experimentally on the MAGPIE facility (Lebedev et al. 2004, 2005). The schematic of the experimental configuration (Fig. 1a) consists of a conical array of micron-sized metallic wires driven by a current of 1 MA rising to its peak value in 240 ns. The basic mechanism of plasma formation in wire arrays is the following: resistive heating rapidly converts the wires into a heterogeneous structure consisting of a cold (< 1 eV) dense, liquid-vapour core surrounded by a relatively hot (10 - 20 eV), low density (∼10$^{17}$ cm$^{-3}$) plasma. Most of the current flows in the low resistivity plasma which undergoes acceleration by the $\boldsymbol{J} \times \boldsymbol{B}$ force towards the array axis. These streams of plasma have characteristic velocities of ∼100 km s$^{-1}$, corresponding to Mach numbers $M \sim 5$. The wire cores act as a reservoir of plasma, replenishing the streams during the entire duration of the experiment (several hundred ns). The converging plasma streams are virtually magnetic field-free and their collision on axis produces a standing conical shock. Although part of the kinetic energy is thermalized at the shock, it is important to note that these streams are not perpendicular to the surface of the shock. The component of the velocity parallel to the shock is continuous across it, and thus the flow is effectively redirected upwards into a jet (Figure 1a). Typical jet velocities attained are ∼ 100 - 200 km s$^{-1}$ and hypersonic jets with *M* > 10 can be produced by this mechanism. The jet collimation and the Mach number depend predominantly on the amount of radiation cooling in the plasma, which can be can be altered by varying wire material (Al, Fe or W). Increasing the atomic number of the wire material increases the rate of energy losses from the plasma, lowers the temperature and leads to the formation of more collimated jets with higher Mach numbers (Ciardi et al. 2002; Lebedev et al. 2002). Typical parameters for a tungsten jet are listed in Table 1.

In the curved jet experiment the wires are made of tungsten, while the cross-wind is produced by a radiatively-ablated plastic foil appropriately placed in the jet propagation region (Ledebev et al. 2004, 2005; Ciardi et al. 2008). Typical "wind" velocities of ~ 30 – 50 km s$^{-1}$ can be produced in the laboratory, with the important parameters characterising the interaction in the range $V_{jet}*/V_{wind}* \sim 2 - 4$ and $\rho_{jet}*/\rho_{wind}* \sim 0.1 - 10$.

In Figure 1b an experimental XUV image of a curved jet shows the presence of internal shocks. The scaling to astrophysical jets is performed by taking a jet to wind speed ratio of 4, and a ratio jet to wind density of 10, which are in a similar range of those characteristic of the laboratory interaction. For the astrophysical simulations of the jet and wind we take $V_{jet}* = 100$ km/s, n_jet = 1000 cm$^{-3}$, and $V_{wind}* = 25$ km/s, n$_{wind}$ = 100 cm$^{-3}$. The initial temperature of both media is 5000 K. In the simulations we only follow the time-dependent populations of atomic and ionized hydrogen. Comparison of the astrophysical and laboratory jet simulations has shown that the dynamics of the interaction is similar for the two systems. Concentrating on the astrophysical jets, the simulations have shown the formation of internal shocks in the jet and the development of a "knotty" flow (Ciardi et al 2008). The curved jets are liable to be Rayleigh-Taylor unstable, with the growth of the instability responsible for disrupting the jet and producing a heterogeneous clumpy flow. A three dimensional view of the resulting flow is shown in Figure 2. Beside disrupting the flow, it is clear that the instability promotes the mixing between the InterStellar Medium and the jet,. The development of the Rayleigh Taylor instability is also expected in the laboratory jets; however for the condition currently produced in the experiments, the growth time is of the order of the dynamical time and new experiments will be needed to observe its full development.

## 5- Laser produced slab-jets

In the last ten years a number of laboratory experiments on high-power lasers have been developed to study high-Mach number jets (Shigemori et al. 2000, Logory et al. 2000, Foster et al. 2002). Although laser produced jets have generally required energies in the kJ range, it was recently shown by Kasperczuk et al. 2006 that direct irradiation of a massive metal target with smaller laser energies (less than ~ few hundred J) can produce collimated, jets in the correct scaling regime. (Kasperczuk et al. 2007; Kasperczuk et al. 2008; Kasperczuk et al. 2009; Schaumann et al. 2005)

Here we present simulations of a variant of this jet generation mechanism, which results in the formation of slab-jets with laser energies of ~ 30 J. Beside the low energy required, these slab-jets are interesting because of their reduced geometry, essentially two-dimensional, which makes them easier to diagnose and simulate. In particular they may be useful to develop tests for 2D numerical simulations and to develop experiments aimed at addressing instabilities (e.g. Kelvin-Helmoltz) linked to the propagation of radiatively cooled jets in the ISM.

The experiments were carried out at the PALS laser facility (Jungwirth 2005, Kozlová et al. 2007) with a beam energy in the range ~ 30 J and a pulse duration of 300 ps ($\omega = 1.315$ $\mu$m), irradiating a planar massive target consisting of an iron foil. The laser focal spot consists of two parallel strips stretched ~ 1 mm in the *x*-direction. Their intensities (~ $1.6 \cdot 10^{13}$ W/cm$^2$) along the y-*direction* have nearly Gaussian distributions (FWHM of 100 μm) and the peaks are separated by ~ 400 $\mu$m. A schematic of the experimental configuration is presented in the figure 3. A typical measurement of the XUV emission recorded through an aluminium window on a XUV CCD camera is shown in figure 4, where an elongated jet-like plasma sheet (~ 1 mm) forms along the symmetry plane (*xz*) between two focal spots.

To understand more precisely the jet formation and the subsequent plasma evolution we have performed two-dimensional simulations (y-*z* slab-geometry) using the 3D code GORGON. The laser energy deposition mechanism, at the relatively low irradiation intensities used in this experiment, is modelled assuming inverse bremsstrahlung absorption. The temporal profile of the laser pulse was approximated by a 300 ps FWHM Gaussian function and given an elongated double-peaked spatial distribution as discussed earlier. A sequence of electron isodensity contours are presented in Fig. 6 and show snapshots of the evolution of the system at 5 ns, 10 ns and 20 ns after the initial laser pulse. The jet formation process is essentially due to the collision on the mid-plane of the two expanding plasma plumes which begins at ~ 5 ns. The converging flows form a shock on the mid-plane which serves to redirect the momentum of the jet in the axial direction, in a mechanism similar to the jets produced from laser irradiated conical targets. At early times (~5 ns), the jet temperature is about 10 eV at 600 microns, with a relatively high degree of ionization (~ 6 at 600 microns). Over time the jet undergoes strong radiative cooling which aids the collimation. The variation of the characteristic jet conditions over 20 ns at ~ 1 mm from the target are density $\rho \sim 2\ 10^{-5}$ to $2\ 10^{-3}$ g cm$^{-3}$, temperature $T \sim 7$ to 3 eV, average degree of ionisation $<Z> \sim 4$ to 1, sound speed $c \sim 9$ to 4 km s$^{-1}$, and flow velocity $u \sim 40$ to 20 km s$^{-1}$. The Mach number in the jet reaches has typical values of 5. The comparison with the characteristic conditions found in astrophysical and z-pinch jets are presented in Table 1, which also shows that the dimensionless numbers Re, Pe and $\chi_{cool}$ are in the correct regime for scaling. By modifying the energy and the profile of the two laser focal spots, it is possible to tune the jet velocity and density. Simulations reveal that it might be possible to achieve a Mach number of ~ 30 by increasing the laser energy by a factor of 5, with the same focal spot geometry.

To complete our investigation and to compare with the experimental results, we have computed the radiative emission of the plasma in the XUV range, including recombination (free-bound) radiation in the range 10 to 100 eV. Figure 5 shows the isocontours of the radiative emission integrated along the *x*-direction and integrated in time between 0 and 100 ns. The simulated emission is in good agreement with the experiments.

**6- Shocks in YSO jets.**

A large variety of shocks are associated with the propagation of YSO jets in the ISM. The strongest shocks are found at the head of the jet: the bow shock that accelerates the ambient medium and the Mach disk that decelerates the jets. There are also weaker shocks between the source and the head, identified in observations by bright knots. Shocks observed in YSO jets are strongly cooled by radiation and present a recombination zone in a thin layer after the shock discontinuity (Hartigan 1994). The description in terms of a cooling function is insufficient to follow the evolution of the different species in the flow (Tesileanu et al. 2008) and these NLTE effects are not scalable in laboratory flows. In general the coupling between hydrodynamics and the time evolution of the populations of different species is difficult to handle because of the stiffness of the latter equations. Furthermore, the coupling between hydrodynamics and radiation transport, which is crucial for modelling the radiative shocks at the head of the fastest jets, introduces another complexity. Although these processes may not be fully scalable, experiments are helpful to test the simulation tools and to understand the physics of these shocks, and a number of radiative shocks experiments were conducted on high-energy laser installations (Bozier et al. 1986, Fleury et al. 2002; Bouquet et al. 2004; Gonzalez et al. 2006; Busquet et al. 2007, Reighard et al. 2005 ).

In a recent experiment, performed at PALS (Gonzalez et al. 2006), shock waves of about 60 km/s were launched in Xenon at 0.2 bar in squared cells of glass with inner section of 0.7 x 0.7 $mm^2$ and length of 4 mm. The cells were closed by a composite foil made from polystyren (10 µm) and a gold film (0.5 µm). The main beam at 435 nm (150 J, 0.3 ns) was focused through a PZP and a lens on a spot of diameter of 0.7 mm on the foil. A scheme of the target is shown in figure 7. The ablation of polystyren foil by the laser generates a strong shock, which propagates through the gold foil and the gas. The gold foil, at the interface between the plastic and the xenon gas, aims at preventing the parasitic X ray radiation generated by the coronal plasma of the laser – plastic interaction to pre-heat the gas inside the cell. Due to the high velocity, the shock front is heated at temperature of 10 – 20 eV. The photons generated at the front propagate in the cold upstream gas, which they heat and ionize. This process thus creates a radiative warm precursor, which precedes the shock front (Figure 8).

In 1D description of the shock wave, all the XUV photons emerging from the shock propagate in the gas and are used for the precursor ionization. However, due to the finite section of the shock tube in the experiments, the photons reach the walls of the tube, and are not necessarily re- injected in the gas, because a fraction F of the flux is lost on the walls (by either absorption or transmission at the walls).

As a consequence of the radiative losses at the walls of the shock tube, the shock waves and the ionization fronts are not flat but become curved, leading to two-dimensional effects on the shock front. To illustrate this point, we performed numerical simulations in 2D with the radiation – hydrodynamics eulerian code HERACLES code (Gonzalez et al. 2007). The shock is driven by a cold piston at a constant velocity of 60 km s$^{-1}$ in xenon at 0.1 bar. When the radiation flux reaches the walls of the cell, a fraction of the flux (F) is lost, while the rest (R=1-F) is reflected secularly into the cell (which is a simplification of the interaction between the radiation and the warm tube walls). We used realistic opacities (Bar Shalom et al. 1989) and equation of state (Michaut et al. 2004) and different conditions for the radiation losses F at the lateral walls (F=10% resulting to a behaviour close to 1D because almost all photons are re-injected in the cell, and F=60% with stronger losses). These losses have an important effect on the shock temperature, especially the precursor extension, which decreases from 1.1mm for F : 10% to 0.5 mm for F= 60%, as shown on Fig. 8.

Radiation losses have thus a strong impact on the extension but also on the dynamics of the radiative precursor (Gonzalez et al. 2009). In Gonzalez et al. 2007, the radiation losses fraction F at the glass tube walls was deduced from the dynamics of the ionisation front, measured by shadowgraphy (fig 7 of this paper) and estimated to F= 60%.

In these strong radiative shocks, the transverse optical depth $\tau$, which tunes the lateral losses and the structure of the precursor, varies form large values (> 10 in the shocked part), intermediate values in the precursor ( ~1) and again large values in the cold gas before the ionisation front. (Gonzalez et al. 2006). It is expected that radiation losses F will have a strong impact on the monochromatic radiative flux, as a consequence of the different sizes of the precursors and of the different temperatures reached in the shock. The influence of the extension of the precursor is illustrated on Fig. 9, which shows the radiation flux emerging at 30 ns from a small hole on the axis, placed at the rear face of a 6 mm tube or at 1.5 mm from the end of the tube (Fig. 8), with an angle of view parallel to the tube. These computations have been using the plasma conditions at the canal center obtained by HERACLES 2D for F = 10 % and F = 60%, at 30 ns, with using a 1D LTE radiative transfer code. This code used monochromatic opacities obtained with hydrogenic model, as described in Michaut et al. 2004. Whereas more adapted mochromatic opacities would be necessary for a spectroscopic analysis of the radiation, this numerical study shows the influence of the losses on the flux through the modification of the precursor structure and the complex variation of the flux within the precursor.

# 7- Conclusions

We have discussed laboratory experiments to study phenomena relevant to stellar jets. These include the effects of a cross-wind on the propagation of jets, which may lead to turbulence, and the formation of knots in curved YSO jets. We have also presented new results related to 2D slab-jet experiments, which may be useful for code validation, and to study jet instability in reduced simpler geometries. Finally, we have presented new results on the radiative properties of experimental radiative shocks, illustrating the limitations of scaling for these shocks in real gases.

# 8- Acknowledmengts:

We acknowledge financial supports from the Access to Research Infrastructures activity in the Sixth Framework Programme of the EU (contract RII3-CT-2003-506350 Laserlab Europe), from the European RTN JETSET (contract MRTN-CT-2004 005592) and from CNRS (PICS 4343). TL gratefully acknowledges financial support from Observatoire de Paris during the past several years. M.G. acknowledges the financial support provided by the Spanish Ministry of Science and Innovation through the Juan de la Cierva grant.

# 8- References

BALLY, J., REIPURTH, B. (2001), Irradiated Herbig-Haro jets in the Orion Nebula and near NGC 1333, *Astroph. J.*, **546**, 299-323.

BAR-SHALOM, A., SHVARTS, D., OREG, J., GOLDSTEIN, W. H., ZIGLER, A. (1989) Super-transition-arrays-A model for the spectral analysis of hot dense plasma, *PRA* 40, 3183- 3193

BIEGING, J. H., COHEN, M., SCHWARTZ, P. R. (1984), VLA observations of T Tauri stars. II - A luminosity-limited survey of Taurus-Auriga, *Astroph. J.*, 282, 699-708.

BOUQUET, S., STEHLÉ, C., KOENIG, M., CHIÈZE, J.P., BENUZZI-MOUNAIX, A., BATANI, D., LEYGNAC, S., FLEURY, X., MERDJI, H., MICHAUT, C., THAIS, F., GRANDJOUAN, N., HALL, T., HENRY, E., MALKA, V., LAFON, J.P.J (2004), Observations of laser driven supercritical radiative shock precursors, *Phys. Rev. Let.* **92** 5001.

BOZIER, J. C., THIELL, G., LEBRETON, J. P., AZRA, S., DECROISETTE, M., SCHIRMANN, D., (1986). Experimental-observation of a radiative wave generated in xenon by a laser-driven supercritical shock, *Phys. Rev. Let.,* **57**, 1304.

CABRIT, S., (2002) Constraints on Accretion-Ejection Structures in Young Stars, in *star formation and the physics of Young Stars* (J. Bouvier J. Zahn J.P. Eds), EDP, p. 147-182

CABRIT, S., (2007) : the need for MHD collimation and Acceleration processes, in *Jets from Young Stars*, Lecture Notes in Physics, **723** (Ferreira, J., Dougados, C., Whelean, E., Eds), Springer, pp 21-50

CASTOR, J.I , (2007), Astrophysical radiation dynamics : the prospect for scaling, *Astrophys. Space Sci.*, **307,** 207

CHITTENDEN, J. P., LEBEDEV, S. V., JENNINGS, C. A., BLAND, S. N., CIARDI, A. (2004), X-ray generation mechanisms in three-dimensional simulations of wire array Z-pinches, *Plasma Physics and Controlled Fusion*, **46**, B457-476

CIARDI, A., AMPLEFORD, D. J., LEBEDEV, S. V., STEHLÉ, C. (2008), Curved Herbig-Haro Jets: Simulations and Experiments, *Astroph. J.*, **678,** 968-973.

CIARDI, A., LEBEDEV, S. V., FRANK, A., BLACKMAN, E. G., CHITTENDEN, J. P.,

JENNINGS, C. J., AMPLEFORD, D. J., BLAND, S. N., BOTT, S. C., RAPLEY, J., HAL,L G. N., SUZUKI-VIDAL, F. A., MAROCCHINO, A., LERY, T., STEHLE, C., The evolution of magnetic tower jets in the laboratory (2007), *Physics of Plasmas,* **14**, 056501-056501-10

DALGARNO, A., MCCRAY, R. A. (1972), Heating and Ionization of HI Regions, *Ann. Rev. Astr. Astroph.,* **10,** p.375

FARLEY, D. R., ESTABROOK, K. G., GLENDINNING, S. G., GLENZER, S. H., REMINGTON, B. A., SHIGEMORI, K., STONE, J. M., WALLACE, R. J., ZIMMERMAN, G. B., HARTE J. A. (1999), Radiative Jet Experiments of Astrophysical Interest Using Intense Lasers , *PRL*, **83**, 1982-1985

FLEURY, X., BOUQUET, S., STEHLÉ, C., KOENIG, M., BATANI, D., BENUZZI-MOUNAIX, A., CHIÈZE, J.P., GRANDJOUAN, N., GRENIER, J., HALL, T. ,HENRY, E., LAFON, J.P.J, LEYGNAC, S., MALKA, V., MARCHET, B., MERDJI, H., MICHAUT, C., THAIS, F., (2002), A laser experiment for studying radiative shocks in astrophysics, *Laser & Particle Beams,* **20,** 263.

GONZÁLEZ, M., AUDIT, E., HUYNH, P. (2007). HERACLES: a three dimensional radiation hydrodynamics code,  *A&A,* **464**, 429-435

GONZÁLEZ, M., STEHLÉ, C. , AUDIT, E. , BUSQUET, M., RUS,  B., THAIS, F., ACEF, O. , BARROSO, P., BAR-SHALOM, A., BAUDUIN, D., KOZLOVA, M., LERY, T., MADOURI, A., MOCEK, T., POLAN, J. (2006) J. Astrophysical radiative shocks : from modelling to laboratory experiments,*Laser Particle Beams*, **24**, 535-545

GONZÁLEZ, M., AUDIT, E., STEHLÉ, C. (2009) 2D numerical study of the radiation influence on shock structure relevant to laboratory astrophysics, *A&A ,* **497**, 27-34

HARTIGAN ,P., RAYMOND, J. (1993), The formation and evolution of shocks in stellar jets from a variable wind, *Astroph. J. ,* **409,** 705-719.

HARTIGAN, P. (1994), Low-Excitation Herbig_Haro Objects, in *Stellar and Circumstellar Astrophysics,* Astron. Soc. Pac. Conf. Proc. (Wallerstein G. & Noriega-Crespo A., Eds), **57,** p.95-104

HARTIGAN, P., FRANK, A., VARNIÉRE, P., BLACKMAN, E. G., (2007), Magnetic Fields in Stellar Jets, *Astroph.J.*, **661**,910

JONES, B. F., HERBIG, G. H. (1979), Proper motions of T Tauri variables and other stars associated with the Taurus-Auriga dark clouds, *Astron. J.*, **84**, 1872-1889

JUNGWIRTH, K., (2005), Recent highlights of the PALS research program, *Laser and Particle Beams*, **23**, no. 2, pp. 177-82.

KASPERCZUK, A., PISARCZYK, T., BORODZIUK, S., ULLSCHMIED, J., KROUSKY, E., MASEK, K., ROHLENA, K., SKALA, J., HORA, H. (2006), Stable dense plasma jets produced at laser power densities around $10^{14}$ W/cm$^2$, *Phys. of Plasmas*, **13**, 062704-062704-8

KASPERCZUK, A, PISARCZYK, T, BORODZIUK, S, ULLSCHMIED, J, KROUSKY, E, MASEK, K, PFEIFER, M, ROHLENA, K, SKALA, J & PISARCZYK, P (2007), Interferometric investigations of influence of target irradiation on the parameters of laser-produced plasma jet', *Laser and Particle Beams*, vol. 25, no. 3, pp. 425-33.

KASPERCZUK, A, PISARCZYK, T, KALAL, M, MARTINKOVA, M, ULLSCHMIED, J, KROUSKY, E, MASEK, K, PFEIFER, M, ROHLENA, K, SKALA, J & PISARCZYK, P (2008), PALS laser energy transfer into solid targets and its dependence on the lens focal point position with respect to the target surface, *Laser and Particle Beams*, vol. 26, no. 2, pp. 189-96.

KASPERCZUK, A, PISARCZYK, T, NICOLAI, PH, STENZ, CH, TIKHONCHUK, V, KALAL, M, ULLSCHMIED, J, KROUSKY, E, MASEK, K, PFEIFER, M, ROHLENA, K, SKALA, J, KLIR, D, KRAVARIK, J, KUBES, P & PISARCZYK, P (2009), Investigations of plasma jet interaction with ambient gases by multi-frame interferometric and X-ray pinhole camera systems, *Laser and Particle Beams*, vol. 27, no. 1, pp. 115-22.

KOZLOVÁ, M., RUS, B., MOCEK,T., POLAN, J., HOMER, P., STUPKA, M., FAJARDO, M., DE LAZZARI, D., ZEITOUN, P. (2007) : Development of Plasma X-ray Amplifiers Based on Solid Targets for the Injector-Amplifier Scheme, In *X-Ray Lasers 2006, Springer Proceedings in Physics* **115** (Ickles V. Anulewicz A. Eds) pp 121-130

LEBEDEV, S.V., CIARDI, A., AMPLEFORD, D.J., BLAND, S.N., BOTT, S.C., CHITTENDEN, J.P., HALL, G.N., RAPLEY, J., JENNINGS, C., SHERLOCK, M., FRANK, A., BLACKMAN, E.G., (2005) : Production of radiatively cooled hypersonic plasma jets and links to astrophysical jets, Plasma *Phys. Control. Fusion*, **47**, B465–B479

LEBEDEV, S. V., AMPLEFORD, D., CIARDI, A., BLAND, S. N., CHITTENDEN, J. P., HAINES, M. G., FRANK, A., BLACKMAN, E. G., CUNNINGHAM, A. (2004), Jet Deflection via Crosswinds: Laboratory Astrophysical Studies, *Astroph. J.* **616** Issue 2 pp. 988-997

LEBEDEV, S. V., CHITTENDEN, J. P., BEG, F. N., BLAND, S. N., CIARDI, A., AMPLEFORD, D., HUGHES, S., HAINES, M. G., FRANK, A., BLACKMAN, E. G., GARDINER, T. (2002) Laboratory Astrophysics and Collimated Stellar Outflows: The Production of Radiatively Cooled Hypersonic Plasma Jets, *Astroph. J.*, **564,** pp. 113-119.

LOGORY, L. M., MILLER, P. L., STRY, P. E. (2000), Nova high-speed jet experiment, *Astroph. J. Sup.*, **127**, 423

MICHAUT, C., STEHLÉ, C., LEYGNAC, S., LANZ, T., BOIREAU, L., (2004), Jump conditions in hypersonic shocks. Quantitative effects of ionic excitation and radiation, *Eur. Phys. J. D*, **28,** 381-392

MIHALAS, D., MIHALAS, B. D. (1984). *Foundation of Radiation Hydrodynamics.*, Oxford University Press.

MUNDT, R., FRIED, J. W. (1983) Jets from young stars *Astroph. J.,* **274,** L83-L86

RAGA, A.C., MELLEMA, G., ARTHUR, S.J., BINETTE, L., FERRUIT, P., STEFFEN, W. (1999), 3D Transfer of the Diffuse Ionizing Radiation in ISM Flows and the Preionization of a Herbig-Haro Working Surface, *Rev. Mex. Astron. Astrof.*, **35,** 123

REIGHARD, A.B., DRAKE, R.P., DANNENBERG, K., PERRY, T.S., ROBEY, H.A., REMINGTON, B.A., WALLACE, R.J., RYUTOV, D.D., GREENOUGH, J., KNAUER, J., BOELHY, T., BOUQUET, S., CALDER, A., ROSNER, R., FRYXELL, B., ARNETT, D., KOENIG, M. (2005), Collapsing radiative shocks in argon gas on the omega laser, ApSS, **298**, Issue 1-2.

REIPURTH, BO RAGA, A. C., HEATHCOTE, S., (1992) Structure and kinematics of the HH 111 jet, *Astroph. J., 392, 145*

REIPURTH, BO; HEATHCOTE, S., MORSE, J., HARTIGAN, P., BALLY, J., (2002), Hubble Space Telescope Images of the HH 34 Jet and Bow Shock: Structure and Proper Motions, *Astron. J.* **123**, 362

RYUTOV, D.D., REMINGTON, B.A., ROBEY, H.F., DRAKE, R.P., (2001), Magneto-hydrodynamic scaling: From astrophysics to the laboratory, *Phys. of Plasmas,* **8,** 1804

RYUTOV, D.D., DRAKE, R.P., KANE, J., LIANG, E., REMINGTON, B.A., WOOD-WASEY,

W.M. (1999), Similarity Criteria for the Laboratory Simulation of Supernova Hydrodynamics, *Astroph. J.,* **518,** 821.

RYUTOV, D.D., DRAKE, R.P., REMINGTON, B.A. (2000), Criteria for Scaled Laboratory Simulations of Astrophysical MHD Phenomena, *Astroph. J. Supp.,* **127,** 465

SALAS, L., CRUZ-GONZALEZ, I., PORRAS, A., (1998) S187 : SCP 1 (H2): A Curved Molecular Hydrogen Outflow, *Astroph. J* .**500**, p.853

SCHAUMANN, G, SCHOLLMEIER, MS, RODRIGUEZ-PRIETO, G, BLAZEVIC, A, BRAMBRINK, E, GEISSEL, M, KOROSTIY, S, PIRZADEH, P, ROTH, M, ROSMEJ, FB, FAENOV, AY, PIKUZ, TA, TSIGUTKIN, K, MARON, Y, TAHIR, NA & HOFFMANN, DHH (2005), High energy heavy ion jets emerging from laser plasma generated by long pulse laser beams from the NHELIX laser system at GSI, *Laser and Particle Beams*, vol. 23, no. 4, pp. 503-12.

SHIGEMORI, K., KODAMA, R., FARLEY, D. R., KOASE, T., ESTABROOK, K. G. , REMINGTON , B. A., RYUTOV, D. D., OCHI, Y., AZECHI, H., STONE, J., TURNER, N., (2000) experiments on radiative collapse in laser-produced plasmas relevant to astrophysical jets, *PR*, **62**, 8838-8841

SNELL, R. L., LOREN, R. B., PLAMBECK, R. L. (1980), Observations of CO in L1551 - Evidence for stellar wind driven shocks, *Astroph. J.*, **239**, L17-L22.

TEŞILEANU, O., MIGNONE A., MASSAGLIA, S., (2008), Simulating radiative astrophysical flows with the PLUTO code: a non-equilibrium multi-species cooling function, *A&A*, **488,** 429-440

ZELDOVICH, Y.B. , RAISER, Y.P. , (1966) *Physics of shock waves and high temperature hydrodynamic phenomena*, New York: Academic

# Figure captions

Figure 1 : Schematic representation of the jet generation using a conical wire array (a), XUV experimental and synthetic images of the bent jet (b) (Ciardi et al 2008)

Figure 2 : 3D isodensity contour (n = 400 cm-3) of a curved astrophysical jet showing the formation of a clumpy flow. The size of the computational box is 2004 x 2004 x 4864 AU.

Figure 3 : Schematic representation of the PALS laser irradiation ($\lambda$=1.315 µm, 30 J, 0.3 ns ) on a massive iron target for laminar jet generation. The peak laser intensity on the target is $1.6 \times 10^{13}$ W/cm$^2$ .

Figure 4 : FFT processed image with low and high frequencies removed of the time integrated XUV emission from the slab-jet. The IR PALS laser pulse comes from the right and the solid iron target is located to the left of the main plasma emission. The positions of the two parallel focal spots are visible in white on the surface of the target (Kozlova et al. 2007).

Figure 5 : Electron isodensity (m$^{-3}$) maps of laser produced jets on PALS at 5, 10 and 20 ns obtained with GORGON.

Figure 6 : Time and space integrated synthetic XUV image of the PALS jet (arbitrary units).

Figure 7 : Typical gaz target design used for radiative shock experiments : the laser ( in green ) impacts the gilt plastic foil. The ablation generates a shock wave which propagates in the z direction. Spectroscopic investigations, for instance from the rear face in XUV may be use to follow the shock radiative signatures.

Figure 8 : Temperature profile ( in eV) of the shock, for two conditions of the walls losses (F=10%

and 60%). The shock propagates at a constant velocity of 60 km/s in a tube of square 0.7 x 0.7 mm$^2$ section (HERACLES simulation), filled with Xenon at 0.1 bar. The temperature in the middle of the tube is reported versus y (along the direction of shock propagation), which is the distance from the position of the maximum of temperature.

Figure 9: Calculated radiative flux along the direction of the tube at 30 ns (in arbitrary units) at different positions of the tube (x=0, i.e. at the end of the tube on the left, and at x=1.5 mm from end of the tube on the right) for two values of the losses at the walls (F=10% in black and F=60% in blue).

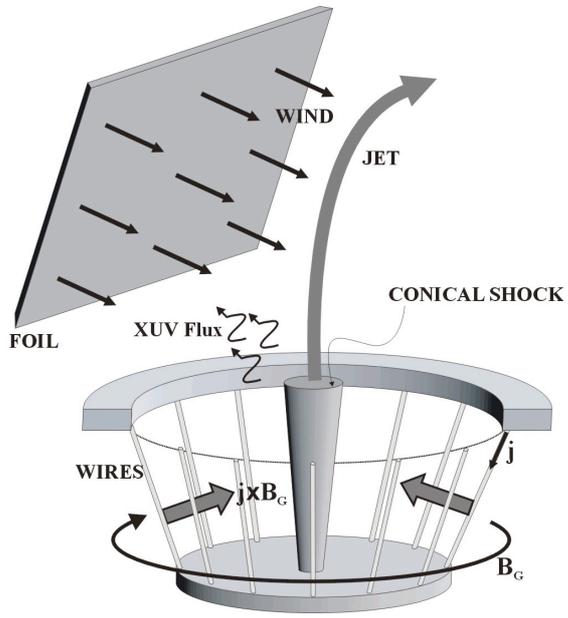 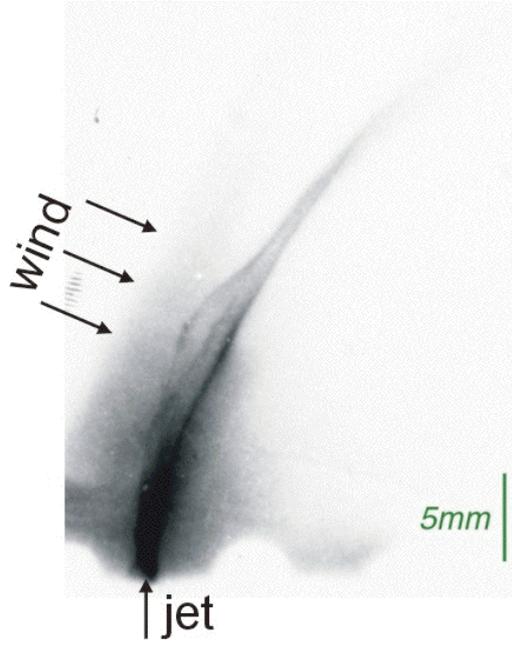

(a) (b)

Figure 1

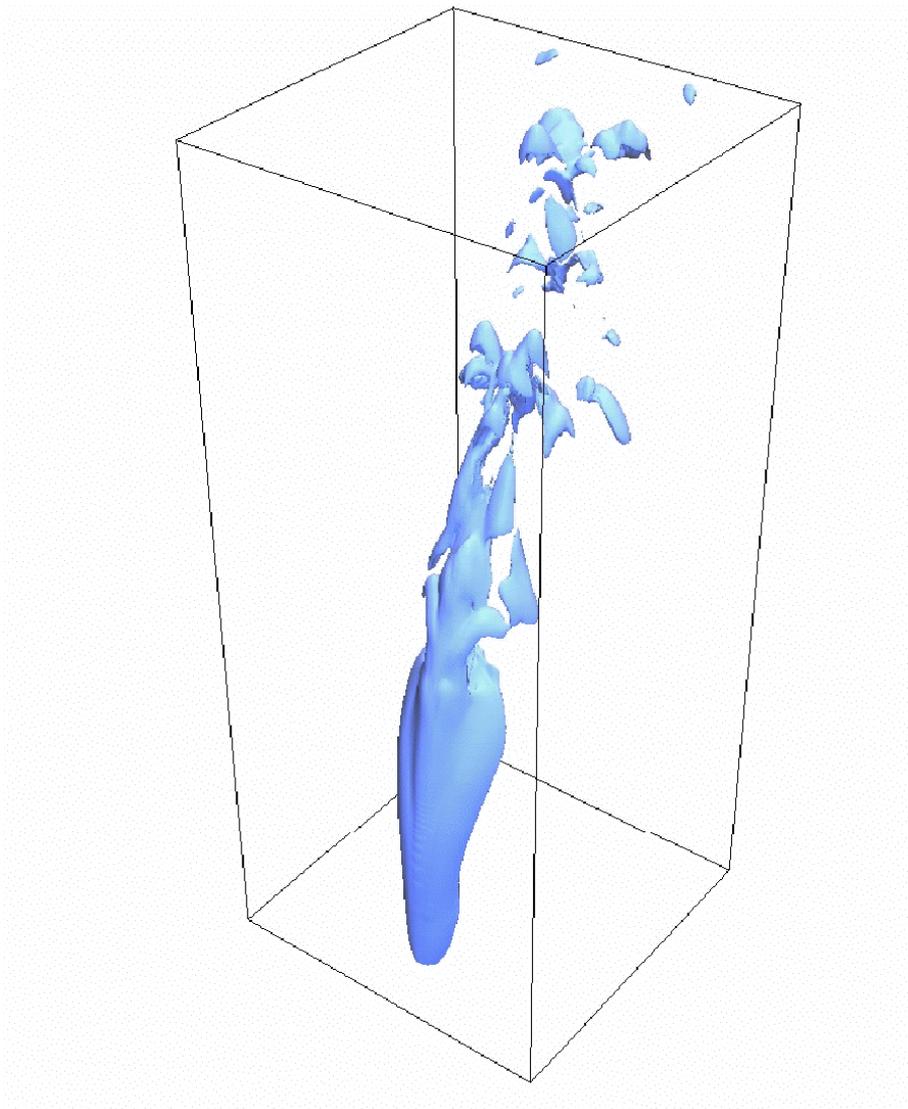

Figure 2

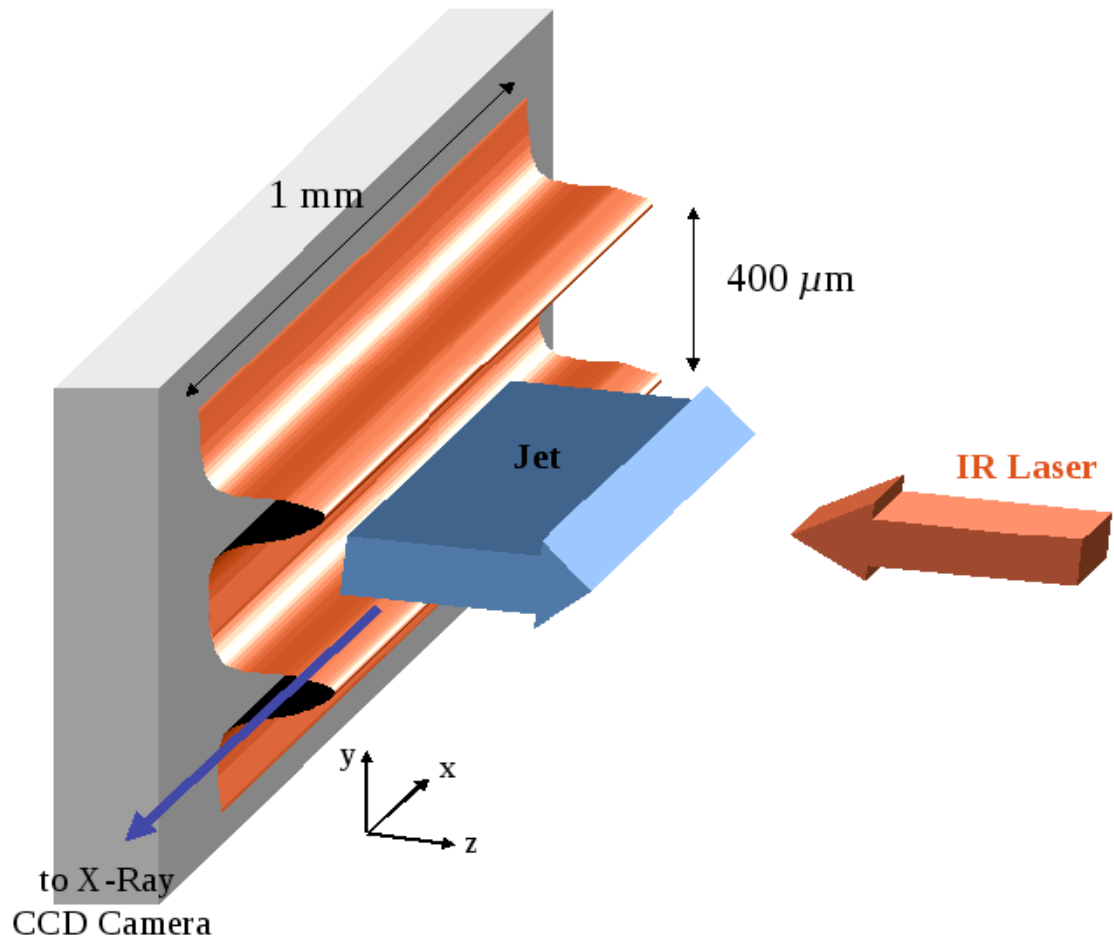

Figure 3

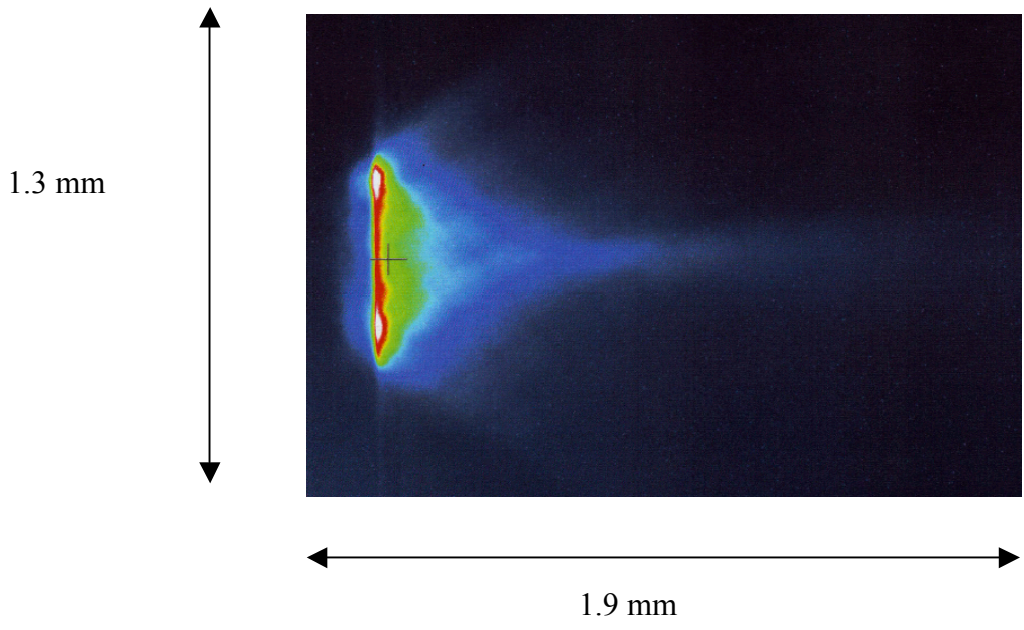

1.3 mm

1.9 mm

Figure 4

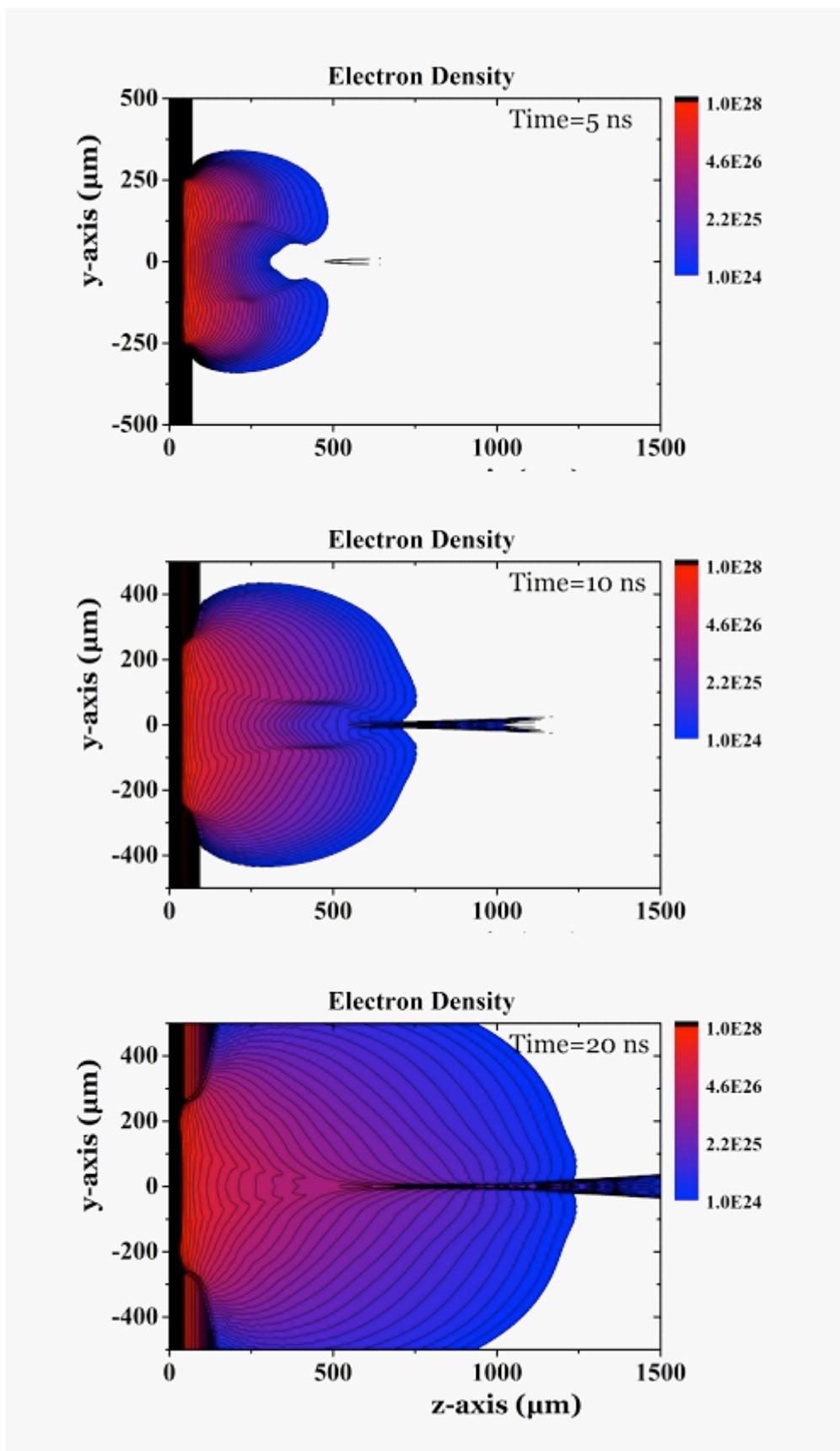

Figure 5

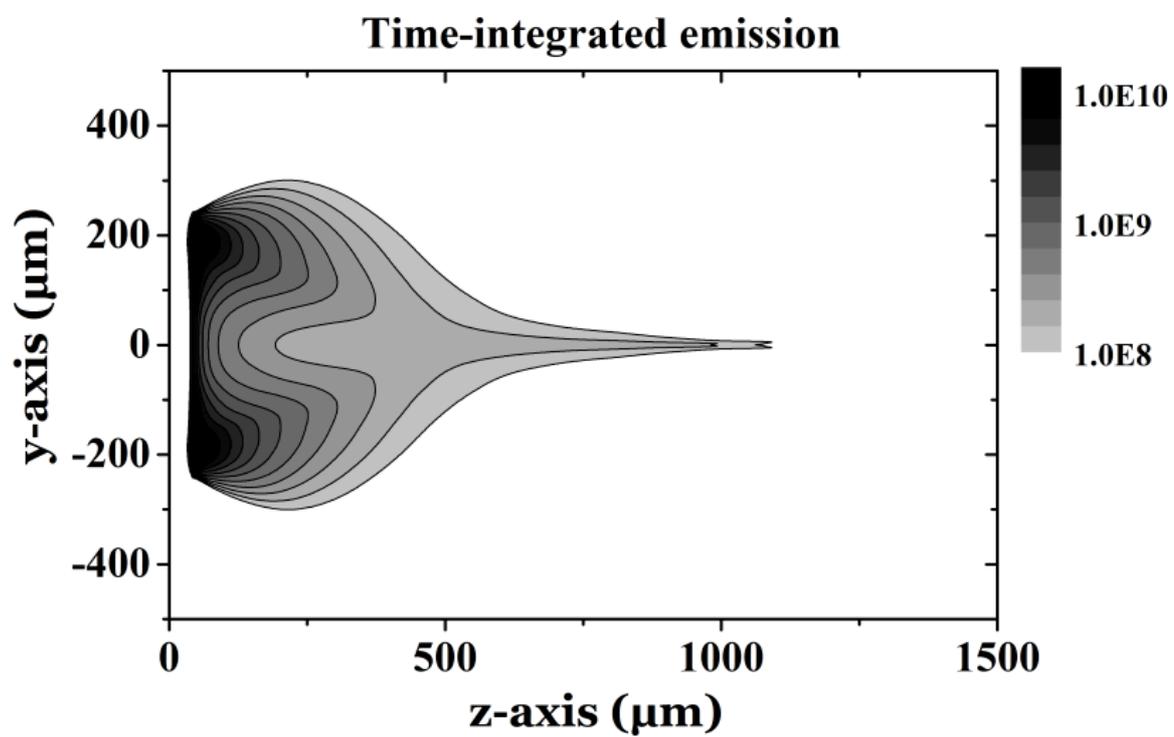

Figure 6

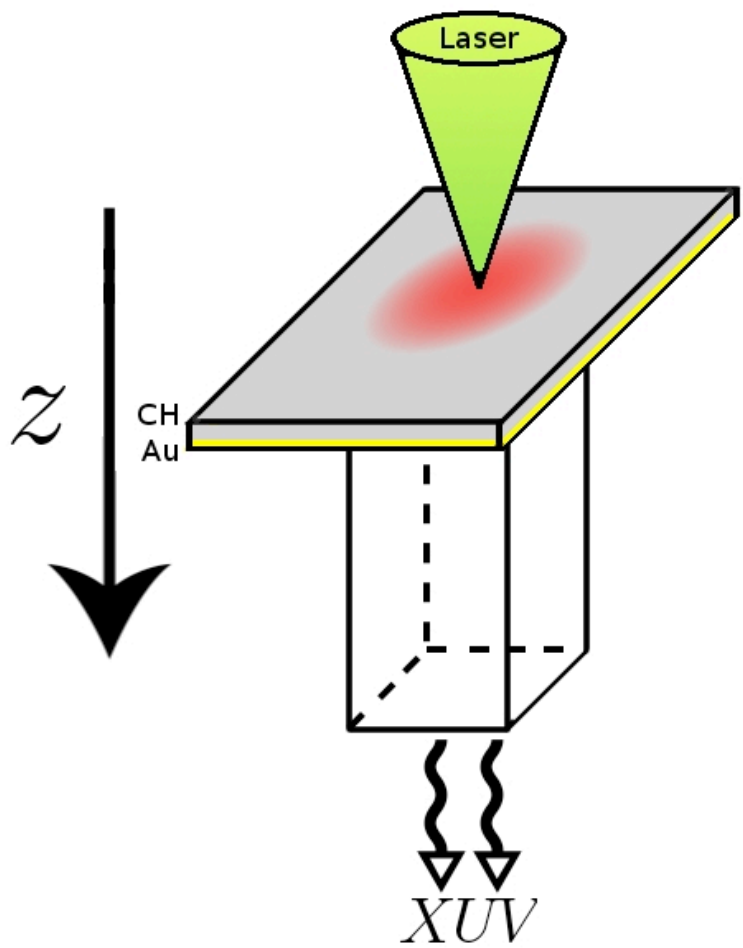

Figure 7

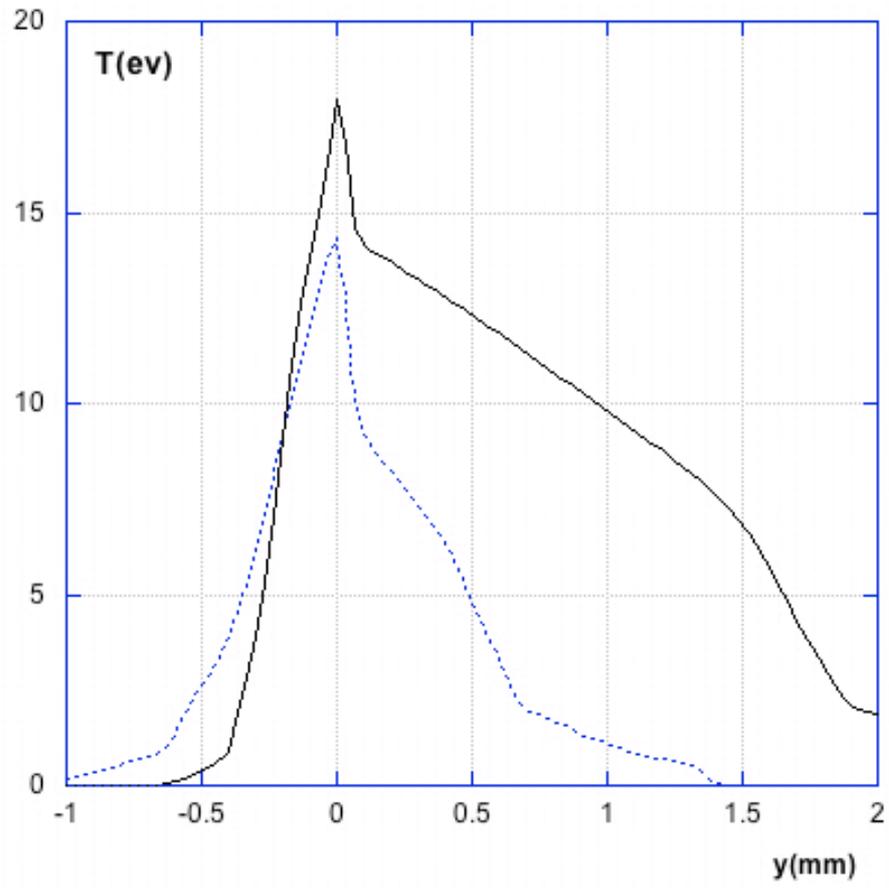

Figure 8

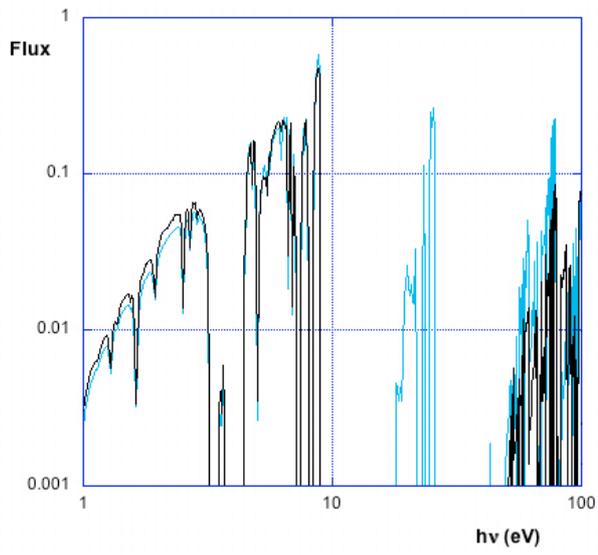 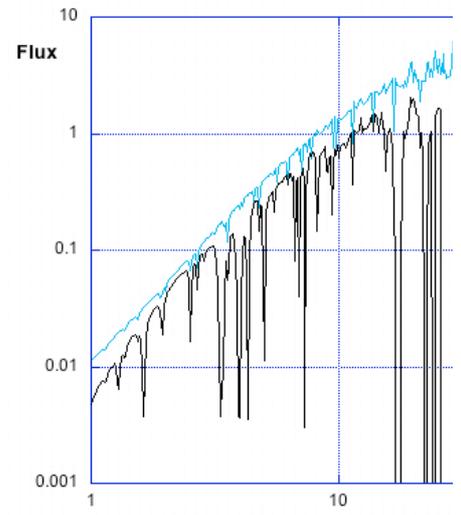

Figure 9

# Table captions

Table 1: Scaling parameters of typical jets : YSO jets, Z –pinch jets and laser jets. Values from YSO jets are taken from Cabrit (Cabrit 2002, 2006). In the case of W experimental jet, the values are taken from Ciardi et al. (2008). The values for the PALS Fe jet are taken from GORGON simulations at 10 ns, on the jet axis and at a distance of 1 mm from the target surface.

| parameter | symbol | YSO jets | Z-pinch jet (W) | PALS jet (Fe) |
|---|---|---|---|---|
| Length (m) | L* | $10^{14}$ | $2\ 10^{-2}$ | $10^{-3}$ |
| Velocity (m/s) | V* | $2\ 10^5$ | $10^5$ | $4\ 10^4$ |
| Temperature (K) | T* | $10^4$ | $3\ 10^4$ | $8\ 10^4$ |
| Ionisation fraction | <Z> | 0.02-0.6 | 5-10 | 4 |
| density (g/cm$^{-3}$) | ρ* | $10^{-19}$ - $10^{-21}$ | $10^{-5}$ | $2\ 10^{-5}$ |
| Mach Number (V/c) | M | 20 - 30 | >20 | 5 |
| Cooling length (m) | $l_{cool}$ | $10^{11}$ | $4\ 10^{-4}$ | $10^{-5}$ |
| Transverse optical depth | τ | <<1 | <<1 | <<1 |
| Reynolds Number | Re | >> $10^5$ | $10^6$ | $10^5$ |
| Peclet Number | Pe | >> $10^4$ | 100 | 8 |

Table 1